\documentclass[a4paper,11pt]{article}
\pdfoutput=1

\usepackage{jheppub}
\usepackage[T1]{fontenc}
\usepackage{hyperref}
\usepackage{graphicx}
\usepackage{amsmath}
\usepackage{epsfig}
\usepackage{subfig}
\usepackage{xcolor}
\usepackage{natbib}
\usepackage{slashed}

\title{NLO fragmentation functions for a quark into a spin-singlet quarkonium: Same flavor case}

\author{Xu-Chang Zheng,}
\author{Xing-Gang Wu}
\author{and Xu-Dong Huang}
\affiliation{Department of Physics, Chongqing University, Chongqing 401331, P.R. China.}
\emailAdd{zhengxc@cqu.edu.cn, wuxg@cqu.edu.cn, hxud@cqu.edu.cn}

\abstract{In the paper, we calculate the fragmentation functions for $c \to \eta_c$ and $b \to \eta_b$ up to next-to-leading-order (NLO) QCD accuracy. The ultraviolet divergences in the real corrections are removed through operator renormalization under the modified minimal subtraction scheme. We then obtain the fragmentation functions $D_{c \to \eta_c}(z,\mu_F)$ and $D_{b \to \eta_b}(z,\mu_F)$ up to NLO QCD accuracy, which are presented as figures and fitting functions. The numerical results show that the NLO corrections are significant. The sensitives of the fragmentation functions to the renormalization scale and the factorization scale are analyzed explicitly.
}


\begin{document}

\maketitle

\bibliographystyle{JHEP}

\section{Introduction}

Since the discovery of the $J/\psi$ in 1974, the heavy quarkonium production has been a focus of theoretical and experimental interest. It is because the production of a heavy quarkonium involves both perturbative and nonperturbative aspects of QCD, and it provides a good platform to study QCD. The most successful effective theory to describe the quarkonium production is the nonrelativistic QCD (NRQCD) effective theory \cite{nrqcd}. Many important quarkonium production processes have been studied up to next-to-leading order (NLO) accuracy under the NRQCD factorization at various colliders \cite{Brambilla:2010cs,Brambilla:2004wf}. However, there are still challenges in understanding the quarkonium production within the NRQCD factorization, such as the $J/\psi$ polarization puzzle \cite{Butenschoen:2012px,Chao:2012iv,Gong:2012ug} and the large differences among various sets of long-distance matrix elements (LDMEs) extracted by several groups, c.f. refs.\cite{Butenschoen:2011yh, Chao:2012iv, Gong:2012ug, Bodwin:2014gia, Bodwin:2015iua}. Therefore, it is important to further study the quarkonium production mechanism.

The production of a quarkonium at high transverse momentum $p_T$ region is simpler than other cases, because the long-distance interactions between the produced quarkonium and initial particles are suppressed. Therefore, in order to explore the quarkonium production mechanism, it is important to study the quarkonium production at high transverse momentum ($p_T$) region.

According to QCD factorization theorem, the production cross section of a hadron ($H$) at high $p_T$ region is dominated by the single parton fragmentation \cite{Collins:1989gx}, i,e.,
\begin{eqnarray}
d\sigma_{A+B \to H(p_T)+X}= \sum_i d \hat{\sigma}_{A+B\to i+X}(p_T/z,\mu_F)
 \otimes D_{i\to H}(z,\mu_F)+{\cal O}(m_H^2/p_T^2),
\label{pqcd-fact}
\end{eqnarray}
where the symbol $\otimes$ represents a convolution integral over the momentum fraction $z$, $d \hat{\sigma}_{A+B\to i+X}(p_T/z,\mu_F)$ are partonic cross sections, $D_{i\to H}(z,\mu_F)$ are fragmentation functions for a parton into a hadron $H$. The sum extends over all species of partons. $\mu_F$ is the factorization scale which separates the energy scales of two parts. The factorization formula (\ref{pqcd-fact}) is also called the leading-power (LP) factorization, since it gives the LP contribution in the expansion in powers of $m_H /p_T$. For the quarkonium production, the next-to-leading power (NLP) contribution can be factorized to double-parton fragmentation \cite{Kang:2011zza, Kang:2011mg, Fleming:2012wy, Fleming:2013qu}.

Unlike the fragmentation functions for light hadrons, the fragmentation functions for quarkonia contain perturbatively calculable information. In fact, the fragmentation functions for the production of quarkonia can be refactorization through the NRQCD factorization, i.e.,
\begin{eqnarray}
D_{i\to H}(z,\mu_F)=\sum_n d_{i\to (Q\bar{Q})[n]}(z,\mu_F) \langle {\cal O}^H(n) \rangle,
\label{frag-nrqcd}
\end{eqnarray}
where $d_{i\to (Q\bar{Q})[n]}(z,\mu_F)$ are short-distance coefficients (SDCs), $\langle {\cal O}^H(n) \rangle$ are NRQCD LDMEs. The SDCs can be calculated perturbatively, while the LDMEs are nonperturbative in nature but can be extracted via global fitting of experimental data or estimated using the Lattice QCD or the QCD potential models.

Most of the fragmentation functions for quarkonia have been calculated up to order $\alpha_s^2$ \cite{Chang:1992bb, Braaten:1993jn, Braaten:1993mp, Braaten:1993rw, Braaten:1994kd, Braaten:1995cj, Chen:1993ii, Yuan:1994hn, Ma:1994zt, Ma:1995ci, Ma:1995vi, Cho:1994gb, Beneke:1995yb, Braaten:2000pc, Hao:2009fa,Sang:2009zz, Jia:2012qx, Bodwin:2014bia, Ma:2013yla, Ma:2015yka, Yang:2019gga}, and a few fragmentation functions for quarkonia have been calculated up to order $\alpha_s^3$ \cite{Artoisenet:2014lpa, Sepahvand:2017gup, Artoisenet:2018dbs, Feng:2018ulg, Zhang:2018mlo, Zheng:2019dfk, Zheng:2019gnb, Feng:2017cjk, Zhang:2020atv, Zheng:2021mqr}.  Among these studies, the fragmentation functions for $q\to \eta_Q$ $(q\neq Q)$, which are of order $\alpha_s^3$, have been calculated in our recent paper \cite{Zheng:2021mqr}. However, the fragmentation functions for $c\to \eta_c$ and $b\to \eta_b$ are only available up to order $\alpha_s^2$. For the precision prediction of the production rate of the $\eta_{c,b}$ at high-energy colliders such as LHC and etc., it is also important to know the fragmentation functions of $c\to \eta_c$ and $b\to \eta_b$ up to order $\alpha_s^3$. In this paper, we will calculate the fragmentation functions $D_{c \to\eta_c}(z,\mu_F)$ and $D_{b \to\eta_b}(z,\mu_F)$ up to NLO QCD accuracy.

The paper is organized as follows. In Sec.II, we present the definition of fragmentation function and the calculation for the LO fragmentation functions. In Sec.III, we sketch the method used in the calculation of the NLO corrections to the fragmentation functions. In Sec.IV, we present the numerical results for the fragmentation functions $D_{c \to\eta_c}(z,\mu_F)$ and $D_{b \to\eta_b}(z,\mu_F)$. Section V is reserved for a summary.

\section{LO fragmentation function}

Before calculating the fragmentation function $D_{Q \to\eta_Q}(z,\mu_F)$, we first present the definition of the fragmentation function for a quark into a hadron. To give the definition for the fragmentation function, it is convenient to work in the light-cone coordinate system. In this coordinate system, a $d$-dimensional vector is expressed as $V^{\mu}=(V^+,V^-,\textbf{V}_{\perp})$, where $V^+=(V^0+V^{d-1})/\sqrt{2}$ and $V^-=(V^0-V^{d-1})/\sqrt{2}$. We adopt a gauge-invariant definition of fragmentation function which was introduced by Collins and Soper in ref.\cite{Collins:1981uw}.

For a quark into a hadron $H$, the fragmentation function is defined as
\begin{eqnarray}
D_{Q\to H}(z)=&&\frac{z^{d-3}}{2\pi}\sum_{X} \int dx^- e^{-iP^+ x^-/z} \frac{1}{N_c} {\rm Tr}_{\rm color}  \frac{1}{4} {\rm Tr}_{\rm Dirac} \left\lbrace \gamma^+ \langle 0 \vert \Psi(0)\right. \nonumber \\
&&\cdot  \bar{{\cal P}}{\rm exp}\left[ig_s \int_{0}^{\infty} dy^- A_a^+(0^+,y^-,{\bf 0}_\perp)t_a^T \right]\vert H(P^+,{\bf 0}_\perp)+X \rangle \nonumber \\
&&\left. \cdot \langle H(P^+,{\bf 0}_\perp)+X\vert {\cal P} {\rm exp}\left[-ig_s \int_{x^-}^{\infty} dy^- A_a^+(0^+,y^-,{\bf 0}_\perp)t_a^T \right] \bar{\Psi}(x)\vert 0\rangle\right\rbrace,
\label{defrag1}
\end{eqnarray}
where $\Psi$ is the initial quark field, $A_a^{\mu}$ is the gluon field, $t_a$ is the color matrix, and ${\cal P}$ indicates the path ordering. The longitudinal momentum fraction is defined as $z\equiv P^+/K^+$, and $K$ is the momentum of the initial quark. This definition of the fragmentation function is carried out in a reference frame where the transverse momentum of the produced hadron $H$ vanishes. It is convenient to introduce a lightlike vector, and in the reference frame, we have $n^{\mu}=(0,1,\textbf{0}_{\perp})$. Then, the ``+" component of a vector can be expressed as $V^+=V\cdot n$, and $z$ can be expressed as a Lorentz invariant $z=P\cdot n/K\cdot n$. According to the definition, the Feynman rules for the fragmentation function can be derived, and the Feynman rules can be found in our previous paper \cite{Zheng:2019gnb}.

To obtain the fragmentation function for $Q \to \eta_Q$, we first calculate the fragmentation function for an on-shell $Q\bar{Q}$ pair in $^1S_0^{[1]}$ state. Then the fragmentation function $D_{Q \to \eta_Q}$ can be obtained from $D_{Q \to (Q\bar{Q})[^1S_0^{[1]}]}$ through replacing the LDME $\langle {\cal O}^{(Q\bar{Q})[^1S_0^{[1]}]}(^1S_0^{[1]}) \rangle$ by $\langle {\cal O}^{\eta_Q}(^1S_0^{[1]}) \rangle$.

\begin{figure}[htbp]
\centering
\includegraphics[width=0.80\textwidth]{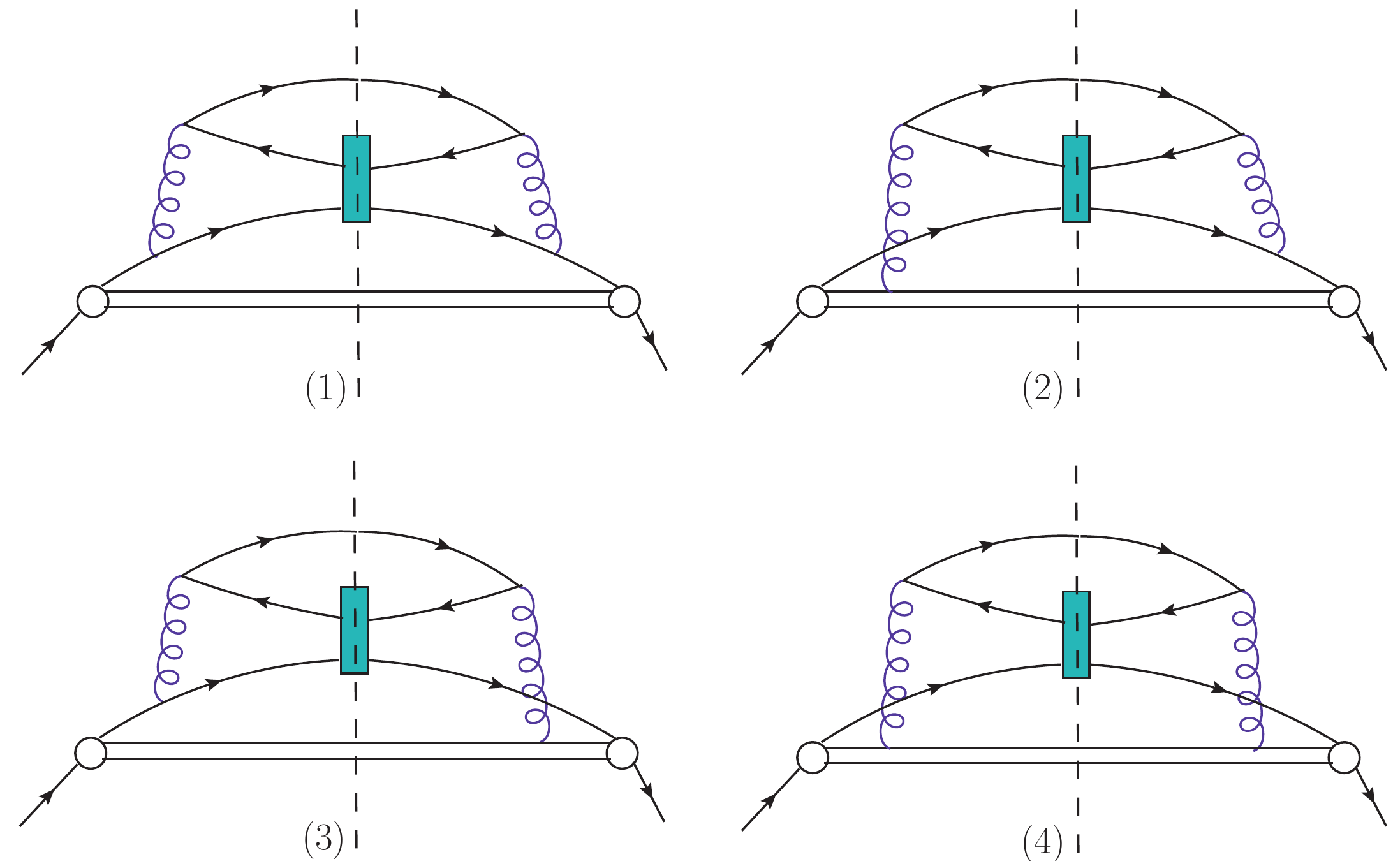}
\caption{The LO cut diagrams for the fragmentation function $D_{Q \to (Q\bar{Q})[^1S_0^{[1]}]}$.
 } \label{feylo}
\end{figure}

Under the Feynman gauge, there are four cut diagrams for $Q(K) \to (Q\bar{Q})[^1S_0^{[1]}](p_1)+Q(p_2)$, as shown in Fig.\ref{feylo}. The squared amplitudes, corresponding to four diagrams in Fig.\ref{feylo}, can be written according to the Feynman rules, i.e,
\begin{eqnarray}
{\cal A}_1=&& {\rm tr}\left[(\slashed{p}_{2}+m_Q)(ig_s\gamma^{\mu}t^a)\Pi_1 \Lambda_1 (ig_s\gamma_{\mu}t^a)  \frac{i}{\slashed{p_1}+\slashed{p_2}-m_Q+i\epsilon} \slashed{n}\frac{-i}{\slashed{p_1}+\slashed{p_2}-m_Q-i\epsilon}\right. \nonumber \\
&& \cdot (-ig_s\gamma^{\nu}t^b)\bar{\Pi}_1\Lambda_1 (-ig_s\gamma_{\nu}t^b) \Big] \frac{-i}{(p_{1}/2+p_2)^2+i\epsilon} \frac{i}{(p_{1}/2+p_2)^2-i\epsilon},\\
{\cal A}_2=&& {\rm tr}\left[(\slashed{p}_{2}+m_Q)(ig_s\gamma^{\mu}t^a)\Pi_1 \Lambda_1 \frac{i}{(-p_1/2-p_2)\cdot n+i\epsilon} (ig_s n_{\mu}t^a)\slashed{n}\frac{-i}{\slashed{p_1}+\slashed{p_2}-m_Q-i\epsilon}\right.\nonumber \\
&& \left. \cdot   (-ig_s\gamma^{\nu}t^b)\bar{\Pi}_1\Lambda_1 (-ig_s\gamma_{\nu}t^b) \right]\frac{-i}{(p_{1}/2+p_2)^2+i\epsilon}\frac{i}{(p_{1}/2+p_2)^2-i\epsilon},\\
{\cal A}_3=&& {\rm tr}\left[(\slashed{p}_{2}+m_Q)(ig_s\gamma^{\mu}t^a)\Pi_1 \Lambda_1 (ig_s\gamma_{\mu}t^a)\frac{i}{\slashed{p_1}+\slashed{p_2}-m_Q+i\epsilon} \slashed{n} (-ig_s n_{\nu}t^b) \right.  \nonumber \\
&& \left. \cdot  \frac{-i}{(-p_1/2-p_2)\cdot n-i\epsilon}\bar{\Pi}_1\Lambda_1 (-ig_s\gamma_{\nu}t^b) \right] \frac{-i}{(p_{1}/2+p_2)^2+i\epsilon}\frac{i}{(p_{1}/2+p_2)^2-i\epsilon} ,\\
{\cal A}_4=&& {\rm tr}\left[(\slashed{p}_{2}+m_Q)(ig_s\gamma^{\mu}t^a)\Pi_1 \Lambda_1 \frac{i}{(-p_1/2-p_2)\cdot n+i\epsilon} (ig_s n_{\mu}t^a)\slashed{n}
 (-ig_s n_{\nu}t^b)  \right. \nonumber \\
&& \cdot \left.  \frac{-i}{(-p_1/2-p_2)\cdot n-i\epsilon} \bar{\Pi}_1\Lambda_1 (-ig_s\gamma_{\nu}t^b) \right]\frac{-i}{(p_{1}/2+p_2)^2+i\epsilon} \frac{i}{(p_{1}/2+p_2)^2-i\epsilon},
\end{eqnarray}
where $\Pi_1$ is the spin projector for the $^1S_0$ state
\begin{eqnarray}
\Pi_1= -\frac{1}{{2\sqrt{2 m_Q}}}\gamma_5 (\slashed{p}_{1} + 2 m_Q),
\end{eqnarray}
and $\bar{\Pi}_1\equiv \gamma^0 \Pi_1^{\dagger} \gamma^0$. $\Lambda_1$ is color-singlet projector
\begin{eqnarray}
\Lambda_1= \frac{\textbf{1}}{\sqrt{3}},
\end{eqnarray}
where $\textbf{1}$ is the unit matrix for the ${\rm SU}_{c}(3)$ group.

Carrying out the color and the Dirac traces, we obtain the expression of the total squared amplitude (${\cal A}_{\rm Born}= \sum_{i=1}^{4}{\cal A}_i$) at LO,
\begin{eqnarray}
{\cal A}_{\rm Born}= \frac{2\,C_F^2\, g_s^4 \, K\cdot n}{ (2-z)^2 m_Q}\sum_{i=1}^3 \frac{a_i\, m_Q^{2(i-1)}}{(s_1-m_Q^2)^{i+1}},\label{aLO}
\end{eqnarray}
where $C_F=4/3$, $s_1=(p_1+p_2)^2$, and
\begin{eqnarray}
a_1=&&(1-z)[(z-2)d-6z+4]^2, \nonumber \\
a_2=&&8z(z-2)[(z-2)d-4z], \nonumber \\
a_3=&&-64(z-2)^2.
\end{eqnarray}

The differential phase space for the fragmentation function at LO is
\begin{eqnarray}
d\phi_{\rm Born} =&& \frac{\theta(p_2^+)dp_2^+}{4\pi p_2^+}\frac{\mu_R^{4-d} d^{d-2}\textbf{p}_{2\perp}}{(2\pi)^{d-2}} 2\pi \delta(K^+-p_1^+-p_2^+),  \label{phase}
\end{eqnarray}
where the $\delta$ function is due to the cut through the eikonal line. The integration over $p_2^+$ can be performed directly using the $\delta$ function. The squared amplitude does not depend on the angles of $\textbf{p}_{2\perp}$, thus the integration over the angles of $\textbf{p}_{2\perp}$ is trivial, and can be performed easily. Then the differential phase space reduces to
\begin{eqnarray}
d\phi_{\rm Born}=&& \frac{z^{-1+\epsilon}(1-z)^{-\epsilon}\mu_R^{2\epsilon}}{2(4\pi)^{1-\epsilon}\Gamma(1-\epsilon)K\cdot n} \left(s_1-\frac{4m_Q^2}{z}-\frac{m_Q^2}{1-z}\right)^{-\epsilon}ds_1.\label{phslo}
\end{eqnarray}
where the relation $s_1=[4m_Q^2/z+m_Q^2/(1-z)+z\, \textbf{p}_{2\perp}^2/(1-z)]$ has been used. The range of $s_1$ is from $[4\,m_Q^2/z+m_Q^2/(1-z)]$ to $+\infty$.

The LO fragmentation function can be obtained through
\begin{eqnarray}
D^{\rm LO}_{Q\to (Q\bar{Q})[^1S_0^{[1]}]}(z)=N_{CS}\int d\phi_{\rm Born} {\cal A}_{\rm Born},\label{fflo}
\end{eqnarray}
where $N_{CS}\equiv z^{d-3}/(8\pi N_c)$ comes from the definition of fragmentation function. Substituting Eqs.(\ref{aLO}) and (\ref{phslo}) into Eq.(\ref{fflo}) and carring out the integral over $s_1$, we obtain
\begin{eqnarray}
D^{\rm LO}_{Q\to (Q\bar{Q})[^1S_0^{[1]}]}(z) =&& \frac{C_F^2\alpha_s^2 z(1-z)(4\pi \mu_R^2)^{\epsilon}\Gamma(1+\epsilon)}{2N_c (2-z)^{4+2\epsilon} m_Q^{3+2\epsilon}}\nonumber \\
&&\cdot \left[a_1+a_2\frac{(1+\epsilon)z(1-z)}{2(2-z)^2} +a_3\frac{(2+\epsilon)(1+\epsilon)z^2(1-z)^2}{6(2-z)^4}\right].
\end{eqnarray}
Setting $d=4$, we obtain
\begin{eqnarray}
D^{\rm LO}_{Q\to (Q\bar{Q})[^1S_0^{[1]}]}(z)=\frac{32\alpha_s^2 z(1-z)^2}{81(2-z)^6 m_Q^3} (3 z^4-8 z^3+8 z^2+48) \label{QQlo}.
\end{eqnarray}
The fragmentation function $D_{Q \to \eta_Q}$ can be obtained from $D_{Q \to (Q\bar{Q})[^1S_0^{[1]}]}$ by multiplying a factor $\langle {\cal O}^{\eta_Q}(^1S_0^{[1]}) \rangle / \langle {\cal O}^{(Q\bar{Q})[^1S_0^{[1]}]}(^1S_0^{[1]}) \rangle \approx \vert R_S(0) \vert^2/(4\pi)$. Then we obtain
\begin{eqnarray}
D^{\rm LO}_{Q\to \eta_Q}(z)=\frac{8\alpha_s^2 z(1-z)^2\vert R_S(0) \vert^2}{81\pi (2-z)^6 m_Q^3} (3 z^4-8 z^3+8 z^2+48) \label{etaQlo},
\end{eqnarray}
which is the same as that obtained in ref.\cite{Braaten:1993mp}.

\section{NLO corrections to the fragmentation function}

At NLO, there are virtual and real corrections contributing to the fragmentation function. In this section, we will briefly introduce the methods used to calculate the virtual and the real corrections.

In the calculation, the package FeynCalc \cite{Mertig:1990an,Shtabovenko:2016sxi} is used to carry out the Dirac and color traces, the packages \$Apart \cite{Feng:2012iq} and Fire \cite{Smirnov:2008iw} are used to do partial fraction and integration-by-part (IBP) reduction. After the IBP reduction, all the one-loop integrals are reduced to master integrals which include the common $A_0$, $B_0$ and $C_0$ functions, and the scalar one-loop integrals with one eikonal propagator. The common $A_0$, $B_0$ and $C_0$ functions are calculated using LoopTools \cite{Hahn:1998yk} numerically, while the scalar integrals with one eikonal propagator are calculated using the method which was introduced in ref.\cite{Artoisenet:2014lpa}.

\subsection{Virtual corrections}

\begin{figure}[htbp]
\centering
\includegraphics[width=0.8\textwidth]{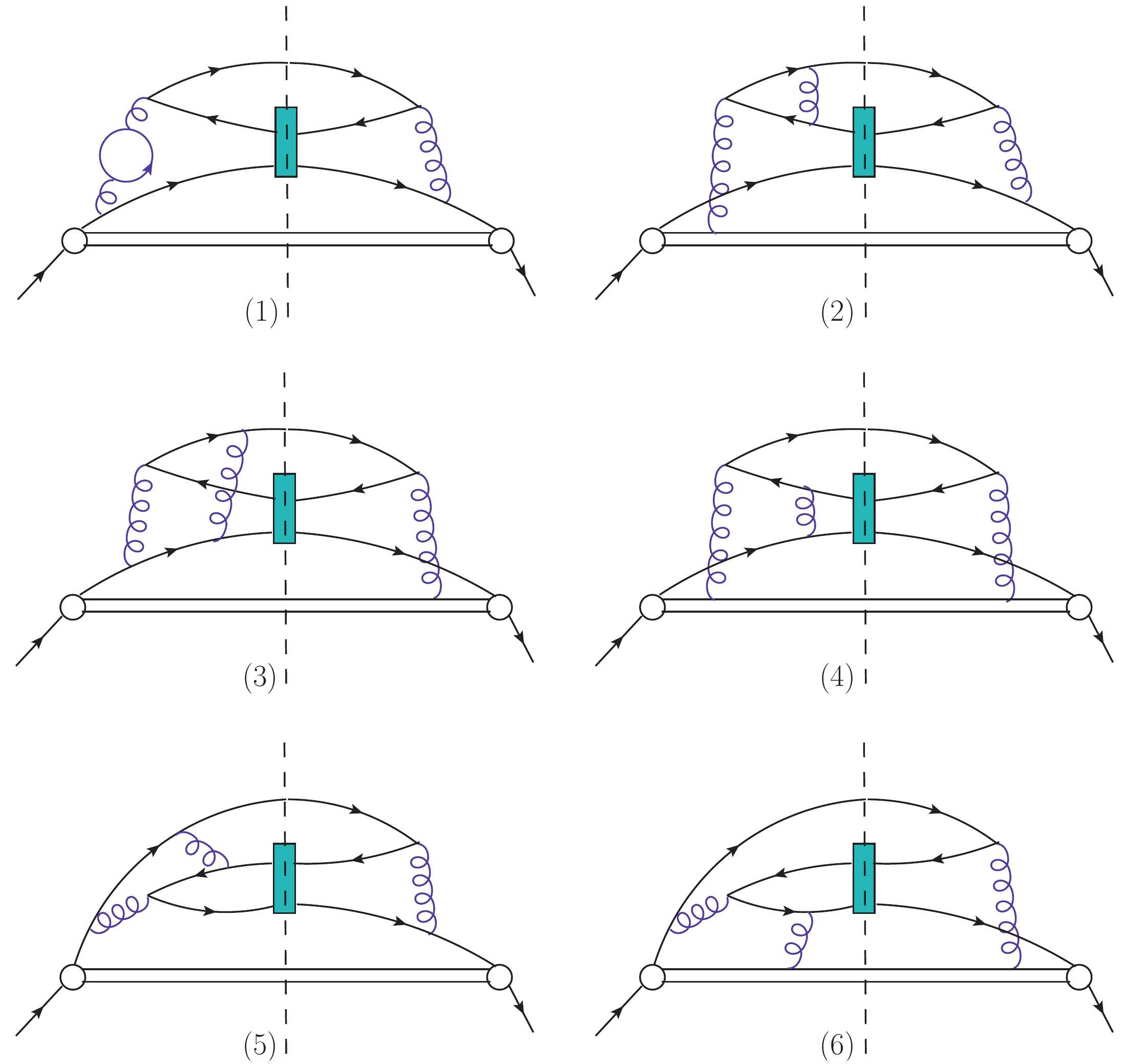}
\caption{Six sample cut diagrams for the virtual corrections to the fragmentation function $D_{Q \to (Q\bar{Q})[^1S_0^{[1]}]}$.
 } \label{feyvir}
\end{figure}

The virtual corrections come from the cut diagrams which contain a loop on either side of the cut line. Six sample cut diagrams for the virtual corrections are presented in Fig.\ref{feyvir}. The virtual corrections can be calculated through
\begin{eqnarray}
D^{\rm virtual}_{Q\to (Q\bar{Q})[^1S_0^{[1]}]}(z)=N_{CS} \int d\phi_{\rm Born} {\cal A}_{\rm virtual},\label{ffvir}
\end{eqnarray}
where ${\cal A}_{\rm virtual}$ denote the squared amplitude for the virtual corrections.

There are ultraviolet (UV) and infrared (IR) divergences in the NLO calculations, dimensional regularization is adopted to regularize these divergences. In dimensional regularization, $\gamma_5$ should be noted. We adopt a practical prescription, which was introduced in ref.\cite{Korner:1991sx}, to handle $\gamma_5$ in dimensional regularization. There are Coulomb divergences in the virtual corrections. Conventionally, these Coulomb divergences are regularized by the relative velocity of the produced $(Q\bar{Q})$ pair, and they should be absorbed into the LDMEs. In this paper, we adopt the threshold expansion method \cite{Beneke:1997zp} to extract the NRQCD SDCs. In this method, we expand the relative momentum $q$ of the produced $(Q\bar{Q})$ pair before carrying out the loop integration. In the leading nonrelativistic approximation, we just set $q=0$ before the loop integration. Then, the Coulomb divergences, which are power divergences, vanish in the calculation.

\subsection{Real corrections}

\begin{figure}[htbp]
\centering
\includegraphics[width=0.8\textwidth]{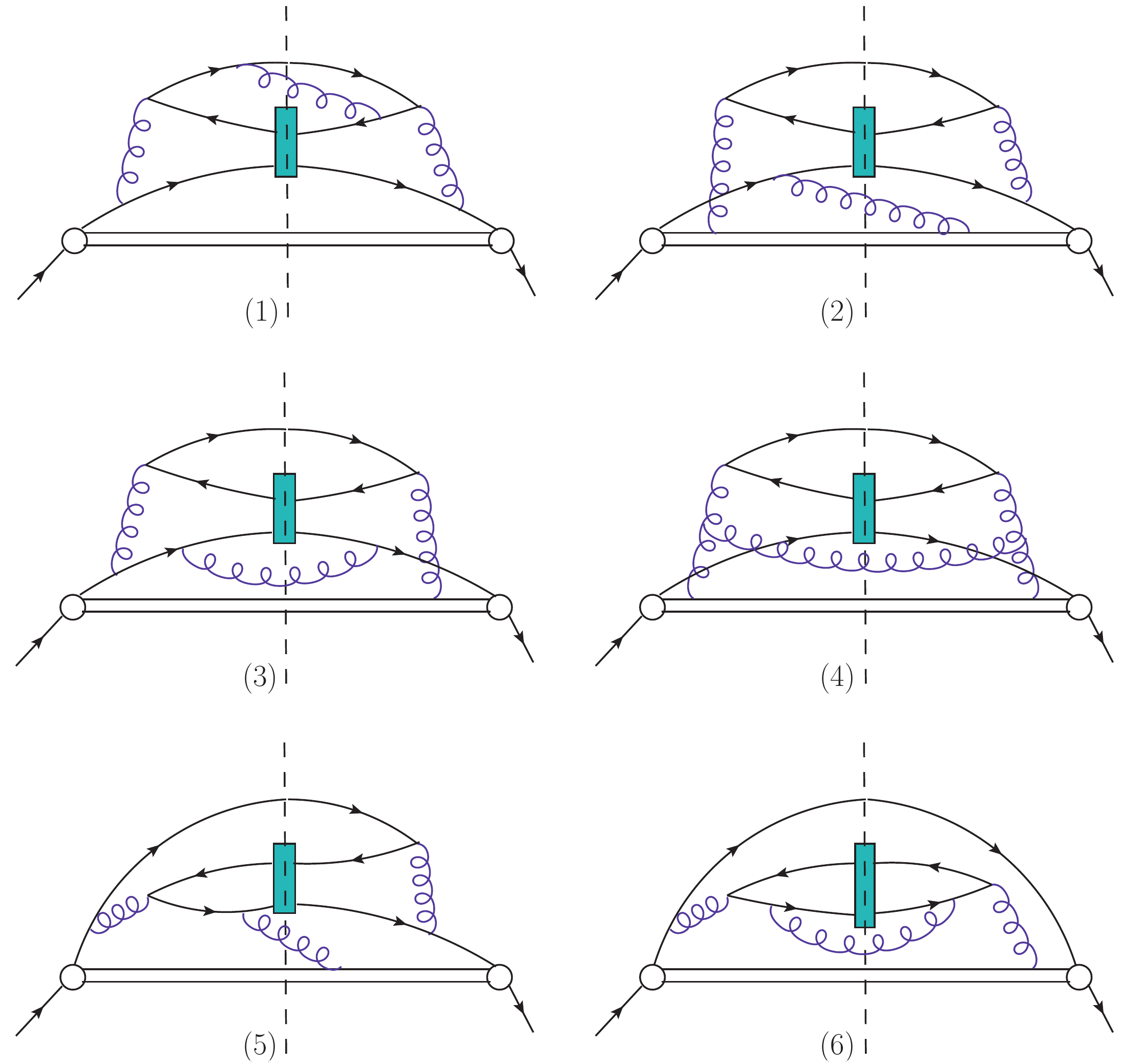}
\caption{Six sample cut diagrams for the real corrections to the fragmentation function $D_{Q \to (Q\bar{Q})[^1S_0^{[1]}]}$.
 } \label{feyreal}
\end{figure}

The real corrections come from the fragmentation process $Q(K) \to (Q\bar{Q})[^1S_0^{[1]}](p_1)+Q(p_2)+g(p_3)$. Six sample cut diagrams are given in Fig.\ref{feyreal}. The real corrections can be obtained through
\begin{eqnarray}
D^{\rm real}_{Q\to (Q\bar{Q})[^1S_0^{[1]}]}(z)=N_{CS} \int d\phi_{\rm real} {\cal A}_{\rm real},\label{ffreal}
\end{eqnarray}
where ${\cal A}_{\rm real}$ denotes the squared amplitude for the real corrections, and the differential phase space for the real corrections is
\begin{eqnarray}
d\phi_{\rm real} = 2\pi \delta \left(K^+-\sum_{i=1}^{3} p_i^+\right) \prod_{i=2,3}\frac{\theta(p_i^+)dp_i^+}{4\pi p_i^+}\frac{\mu_R^{4-d} d^{d-2}\textbf{p}_{i\perp}}{(2\pi)^{d-2}},\label{phase-real}
\end{eqnarray}

There are UV and IR divergences in the real corrections. These divergences come from the phase space integrals and should be regularized by dimensional regularization as that in the virtual corrections. In order to isolate the divergent and the finite terms, we adopt the subtraction method. Under this method, the real corrections are expressed as
\begin{eqnarray}
D^{\rm real}_{Q\to (Q\bar{Q})[^1S_0^{[1]}]}(z)=&& N_{CS} \int d\phi_{\rm real} \left({\cal A}_{\rm real}-{\cal A}_{S}\right) + N_{CS} \int d\phi_{\rm real} {\cal A}_{S},\label{ffsub}
\end{eqnarray}
where ${\cal A}_{S}$ denotes the subtraction term, which has the same singularities as ${\cal A}_{\rm real}$. The first term on the right-hand side of Eq.(\ref{ffsub}) is finite, thus it can be calculated in 4 dimensions directly. The second term on the right-hand side of Eq.(\ref{ffsub}) is divergent, and it should be calculated in $d$ dimensions analytically.

The methods of constructing and integrating the subtraction terms can be found in our previous paper \cite{Zheng:2019gnb}. It should be noted that there are new types of cut-diagrams (e.g, the fifth and the sixth diagrams in Fig.\ref{feyreal}) in the $\eta_Q$ case compared to the $J/\psi$ case. There are additional subtraction terms arise from these new-type cut diagrams, and the integration of these additional subtraction terms can be found in our another paper \cite{Zheng:2021mqr}. With those formulas presented in refs.\cite{Zheng:2019gnb,Zheng:2021mqr}, the real corrections can be calculated directly.

\subsection{Renormalization}

The calculation is carried out with the renormalized fields $\Psi_R$ and $A_R^{\mu}$, the renormalized quark mass $m_Q$, and the renormalized strong coupling constant $g_s$. The relations between the renormalized and the bare quantities are
\begin{eqnarray}
\Psi_0=Z_2^{1/2} \Psi_R, ~~ A_0^{\mu}=Z_3^{1/2} A_R^{\mu},~~ m_{Q0}=Z_m\,m_Q,~~ g_{s0}=Z_g\,g_s,
\end{eqnarray}
where $Z_i \equiv 1+\delta_i$ are renormalization constants. The renormalization constants are fixed by the renormalization scheme. In this paper, the renormalization of the strong coupling constant is carried out in the modified minimal-subtraction scheme ($\overline{\rm MS}$), while the renormalization of the quark field, the gluon field and the quark mass $m_Q$ are carried out in the on-mass-shell scheme (OS). The expressions of the quantities $\delta_i$ are
\begin{eqnarray}
\label{rencont}
 \delta Z^{\overline{\rm MS}}_g&=&- \frac{\beta_0}{2}\frac{\alpha_s}{4\pi}\left[\frac{1}{\epsilon_{UV}}- \gamma_E+ {\rm ln}~(4\pi) \right]\nonumber\\
\delta Z^{\rm OS}_2&=&-C_F \frac{\alpha_s}{4\pi}\left[\frac{1}{\epsilon_{UV}}+ \frac{2}{\epsilon_{IR}}-3~\gamma_E+3~ {\rm ln}\frac{4\pi \mu_R^2}{m^2}+4\right], \nonumber\\
 \delta Z^{\rm OS}_3&=&\frac{\alpha_s}{4\pi}\left[(\beta'_0-2C_A)\left(\frac{1}{\epsilon_{UV}}-\frac{1}{\epsilon_{IR}}\right) \right. \nonumber\\
 &&\left.-\frac{4}{3}T_F \left(\frac{1}{\epsilon_{UV}}-\gamma_E + {\rm ln}\frac{4\pi \mu_R^2}{m_c^2}\right)\right. \nonumber\\
 &&\left.-\frac{4}{3}T_F \left(\frac{1}{\epsilon_{UV}}-\gamma_E + {\rm ln}\frac{4\pi \mu_R^2}{m_b^2}\right)\right], \nonumber\\
 \delta Z^{\rm OS}_m&=&-3~C_F \frac{\alpha_s}{4\pi}\left[\frac{1}{\epsilon_{UV}}- \gamma_E+
 {\rm ln}\frac{4\pi \mu_R^2}{m^2}+\frac{4}{3}\right],
\end{eqnarray}
where $C_F=4/3$, $T_F=1/2$, $\beta_0=11-2\, n_f /3$, $\beta_0'=11-2\, n_{lf} /3$, $n_f$ is the number of the active-quark flavors and $n_{lf}$ is the number of the light-quark flavors. The contributions from these counterterms can be obtained through
\begin{eqnarray}
D^{\rm counter}_{Q\to (Q\bar{Q})[^1S_0^{[1]}]}(z)=N_{CS}\int d\phi_{\rm Born} {\cal A}_{\rm counter},\label{ffct}
\end{eqnarray}
where ${\cal A}_{\rm counter}$ denotes the squared amplitude for the counterterms from the strong coupling constant, the quark field, the gluon field and the quark mass.

The operator used to define the fragmentation function is also need to be renormalized \cite{Mueller:1978xu}. We carry out the operator renormalization in the $\overline{\rm MS}$ scheme, and
\begin{eqnarray}
D^{\rm CT,operator}_{Q\to (Q\bar{Q})[^1S_0^{[1]}]}(z)=&&-\frac{\alpha_s}{2\pi}\left[\frac{1}{\epsilon_{UV}}- \gamma_E+ {\rm ln}~(4\pi)+{\rm ln}\frac{\mu_R^2}{\mu_F^2} \right]\int_z^1 \frac{dy}{y} \nonumber \\
&&\cdot \left[ P_{QQ}(y)D_{Q\to (Q\bar{Q})[^1S_0^{[1]}]}^{\rm LO}(z/y) +P_{gQ}(y)D_{g\to (Q\bar{Q})[^1S_0^{[1]}]}^{\rm LO}(z/y)\right] ,
\end{eqnarray}
where $D_{Q\to (Q\bar{Q})[^1S_0^{[1]}]}^{\rm LO}(z)$ and $D_{g\to (Q\bar{Q})[^1S_0^{[1]}]}^{\rm LO}(z)$ denote the LO fragmentation functions in $d$-dimensional space-time. The splitting functions
\begin{equation}
P_{QQ}(y)=C_F\left[\frac{1+y^2}{(1-y)_+}+\frac{3}{2}\delta(1-y)\right]
\label{lospfun}
\end{equation}
and
\begin{eqnarray}
P_{gQ}(y)=C_F \frac{1+(1-y)^2}{y}.
\end{eqnarray}

\section{Numerical results and discussion}

Summing the contributions from the virtual and real corrections and the counter terms, the UV and IR divergences are canceled, and the finite fragmentation function up to NLO QCD accuracy is obtained. The fragmentation function for the $\eta_Q$ can be obtained from the fragmentation function for the $(Q\bar{Q})[^1S_0^{[1]}]$ pair by multiplying a factor $\langle {\cal O}^{\eta_Q}(^1S_0^{[1]}) \rangle / \langle {\cal O}^{(Q\bar{Q})[^1S_0^{[1]}]}(^1S_0^{[1]}) \rangle \approx \vert R_S(0) \vert^2/(4\pi)$. In doing the numerical calculation, the input parameters are taken as follows:
\begin{eqnarray}
m_c=1.5\,{\rm GeV},~~m_b=4.9\,{\rm GeV},~\vert R_S^{(c\bar{c})}(0)\vert^2=0.810\,{\rm GeV}^{3},~\vert R_S^{(b\bar{b})}(0)\vert^2=6.477\,{\rm GeV}^{3},
\end{eqnarray}
where $R_S^{(c\bar{c})}(0)$ and $R_S^{(b\bar{b})}(0)$ are the radial wave functions at the origin for the $(c\bar{c})$ and $(b\bar{b})$ systems, and the input values for them are taken from the potential-model calculation \cite{Eichten:1995ch}. For the strong coupling constant, we adopt the two-loop running formula, i.e.,
\begin{displaymath}
\alpha_s(\mu_R)=\frac{4\pi}{\beta_0{\rm ln}(\mu_R^2/\Lambda^2_{QCD})}\left[ 1-\frac{\beta_1{\rm ln}\,{\rm ln}(\mu_R^2/\Lambda^2_{QCD})}{\beta_0^2\,{\rm ln}(\mu_R^2/\Lambda^2_{QCD})}\right],
\end{displaymath}
where $\beta_1=102-38\,n_f/3$ is the two-loop coefficient of the QCD $\beta$-function. According to $\alpha_s(m_{_Z})=0.1185$~\cite{Patrignani:2016xqp}, we obtain $\Lambda^{n_f=5}_{\rm QCD}=0.233\,{\rm GeV}$ and $\Lambda^{n_f=4}_{\rm QCD}=0.337\,{\rm GeV}$.

\begin{figure}[htbp]
\centering
\includegraphics[width=0.8\textwidth]{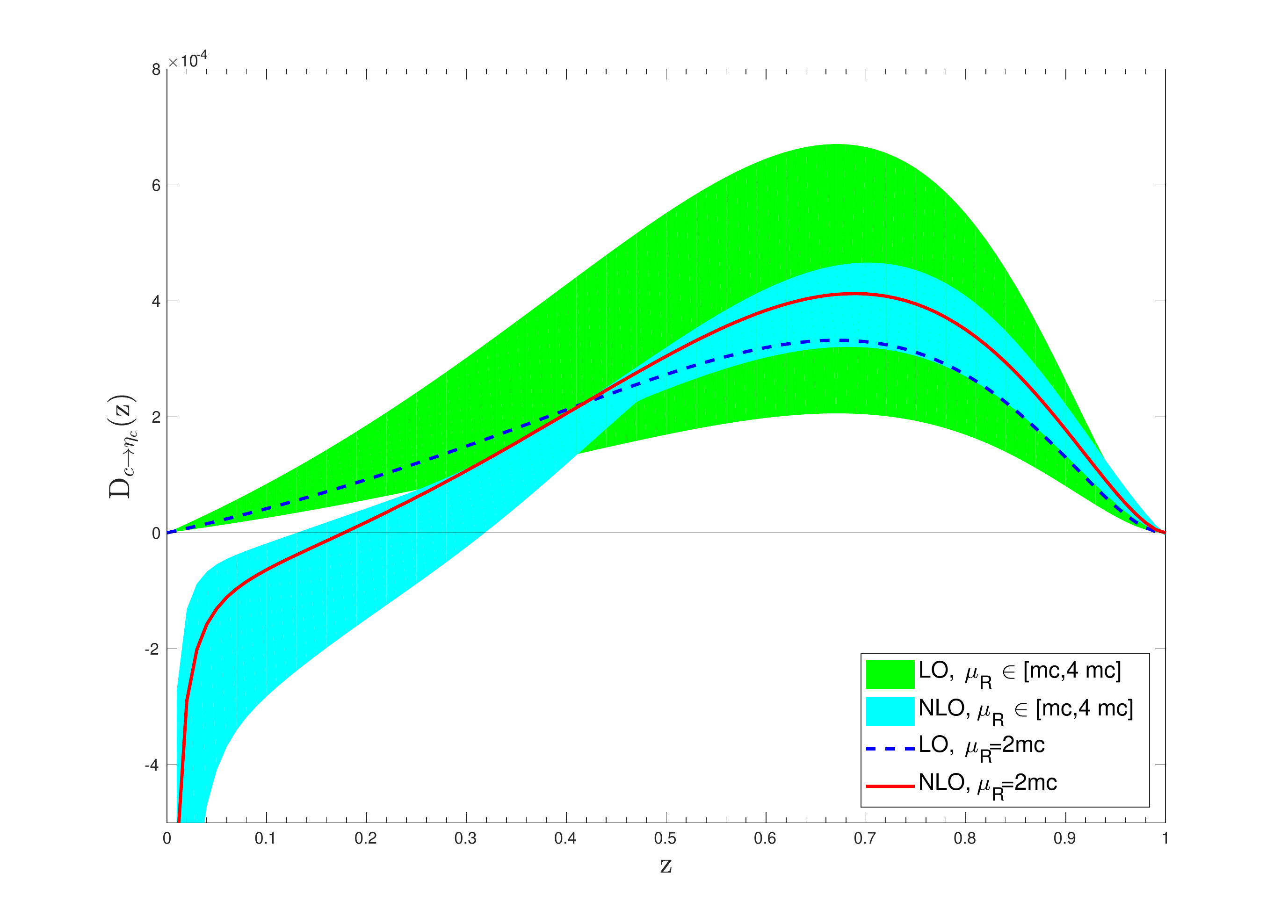}
\caption{The fragmentation function $D_{c\to \eta_c}(z,\mu_{F})$ as a function of $z$ up to LO and NLO accuracy, where $\mu_F=3m_c$ and $\mu_R=2m_c$. The bands are obtained by varying the renormalization scale $\mu_R$ by a factor of 2.} \label{dz-etac-mur}
\end{figure}

The LO and NLO fragmentation functions for $c\to \eta_c$ are shown in Fig.\ref{dz-etac-mur}. Here, the renormalization scale is set as $\mu_R=2m_c$, i.e., the minimal invariant mass of the gluon in the LO fragmentation process; the factorization scale is set as $\mu_F=3m_c$, i.e., the minimal invariant mass of the initial $c$ quark. From the figure, we can see that the effect of the NLO corrections is significant. The fragmentation function is decreased at small $z$ values and increased at large $z$ values after including the NLO corrections. The NLO fragmentation function behaves like $1/z$ as $z \to 0$, while the LO fragmentation function tends to 0 as $z \to 0$. This is because that there are new type cut diagrams (e.g., the sixth diagram in Fig.\ref{feyreal}) contributing to the real correction, and these new type cut diagrams are the same as those of a light quark into the $\eta_c$. As shown in our previous paper \cite{Zheng:2021mqr}, the fragmentation function $D_{q \to \eta_c}(z,\mu_F)$ behaves like $1/z$ as $z \to 0$, then the $1/z$ behavior of the NLO fragmentation function for $c\to \eta_c$ is expected.

The sensitivity of the LO and NLO fragmentation functions to the renormalization scale $\mu_R$ is also shown in Fig.\ref{dz-etac-mur}. The bands in Fig.\ref{dz-etac-mur} are obtained by varying the renormalization scale $\mu_R$ by a factor of 2 around the center value $2 m_c$. From the figure, we can see that the sensitivity of the NLO fragmentation function to $\mu_R$ is decreased at the moderate and the large $z$ values, but increased at the small $z$ values, compared to the LO fragmentation function. The reason for the large sensitivity of the NLO fragmentation function to $\mu_R$ at small $z$ values is that the NLO correction is large compare with the LO contribution at the small $z$ values.

\begin{figure}[htbp]
\centering
\includegraphics[width=0.8\textwidth]{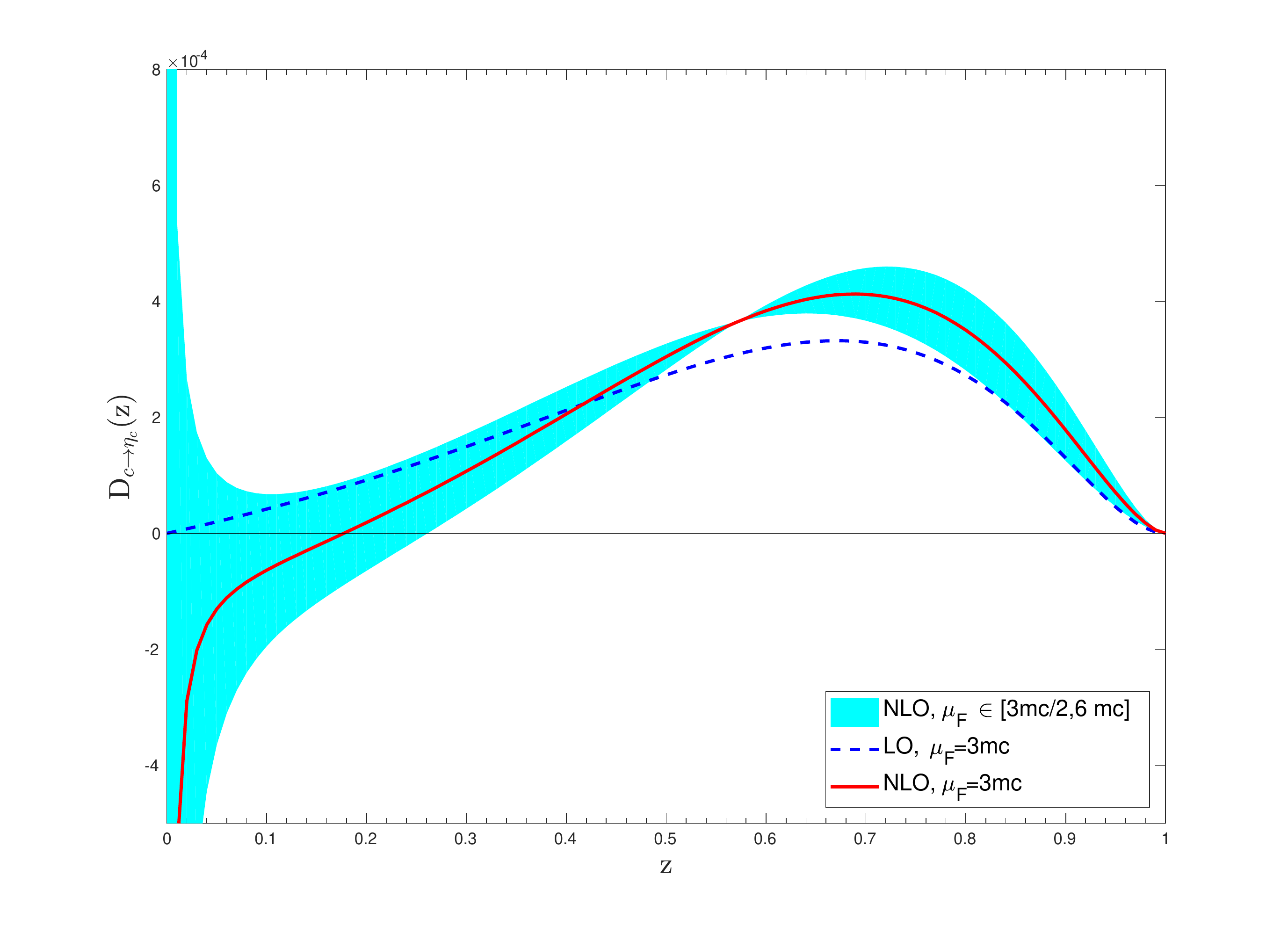}
\caption{The fragmentation function $D_{c\to \eta_c}(z,\mu_{F})$ as a function of $z$ up to LO and NLO accuracy, where $\mu_F=3m_c$ and $\mu_R=2m_c$. The bands are obtained by varying the factorization scale $\mu_F$ by a factor of 2.} \label{dz-etac-muf}
\end{figure}

The sensitivity of the LO and NLO fragmentation functions to the factorization scale $\mu_F$ is shown in Fig.\ref{dz-etac-muf}. The band in Fig.\ref{dz-etac-muf} is obtained by varying the factorization scale $\mu_F$ by a factor of 2 around the center value $3m_c$. From the figure, we can see that the LO fragmentation function is independent of $\mu_F$, the $\mu_F$ dependence of the fragmentation function starts at the NLO. The NLO fragmentation function is very sensitive to $\mu_F$ when $z$ is very small, which is similar to the fragmentation function of $q \to \eta_c$  \cite{Zheng:2021mqr}. It is noted that the dependence of the fragmentation function on $\mu_F$ is not a theoretical uncertainty, because the fragmentation function is always defined at a specified scale $\mu_F$.

The fragmentation function $D_{c\to \eta_c}(z,\mu_{F})$ contains logarithms of $\mu_F/m_Q$, and these logarithms may spoil the the convergence of the perturbative expansion of the fragmentation function when $\mu_F \gg m_c$. These large logarithms can be resummed through solving the Dokshitzer-Gribov-Lipatov-Altarelli-Parisi (DGLAP) evolution equations \cite{dglap1,dglap2,dglap3}, where the fragmentation function with $\mu_F \sim m_c$ is used as the boundary condition.

\begin{figure}[htbp]
\centering
\includegraphics[width=0.8\textwidth]{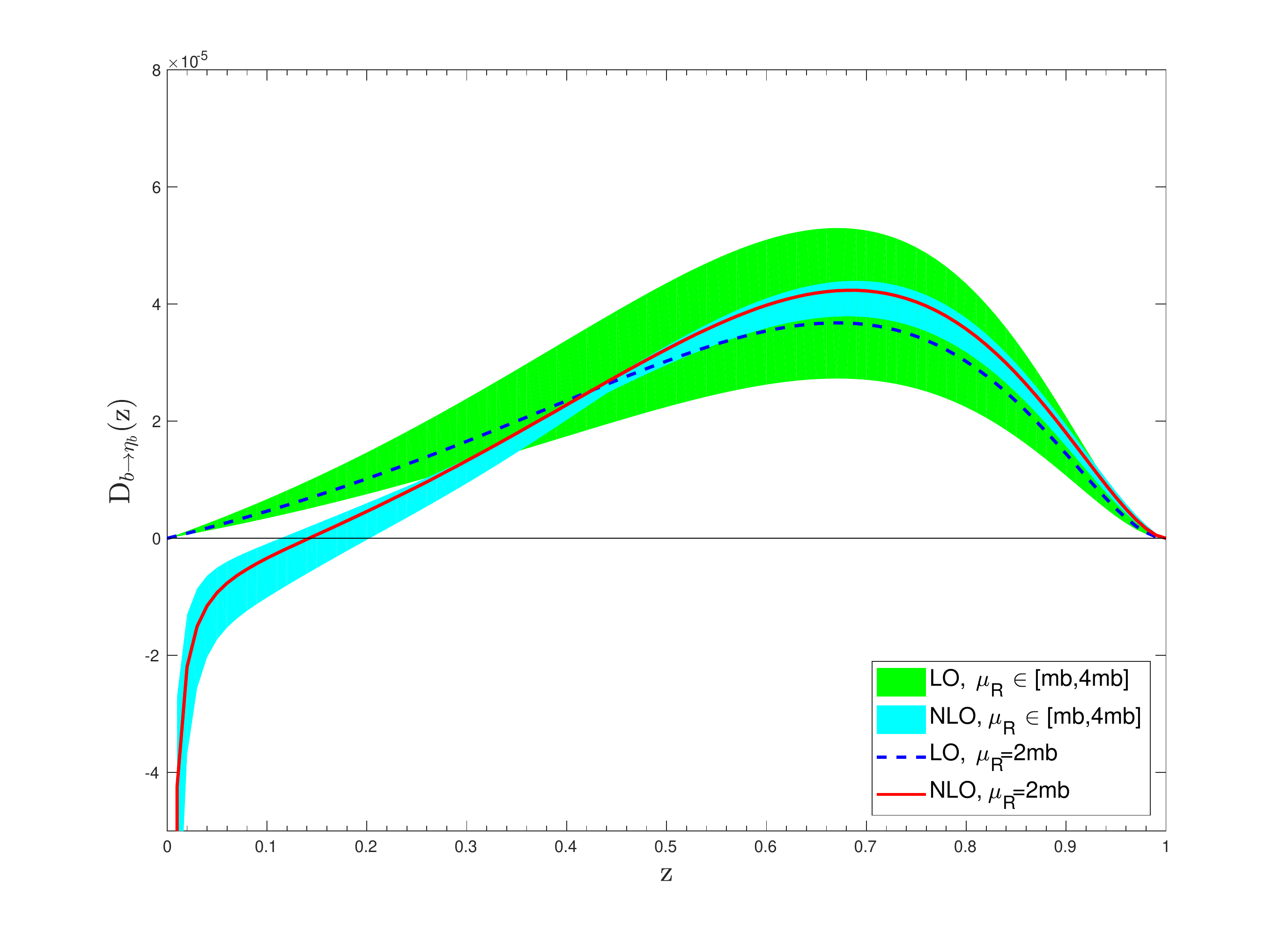}
\caption{The fragmentation function $D_{b\to \eta_b}(z,\mu_{F})$ as a function of $z$ up to LO and NLO accuracy, where $\mu_F=3m_b$ and $\mu_R=2m_b$. The bands are obtained by varying the renormalization scale $\mu_R$ by a factor of 2.} \label{dz-etab-mur}
\end{figure}

\begin{figure}[htbp]
\centering
\includegraphics[width=0.8\textwidth]{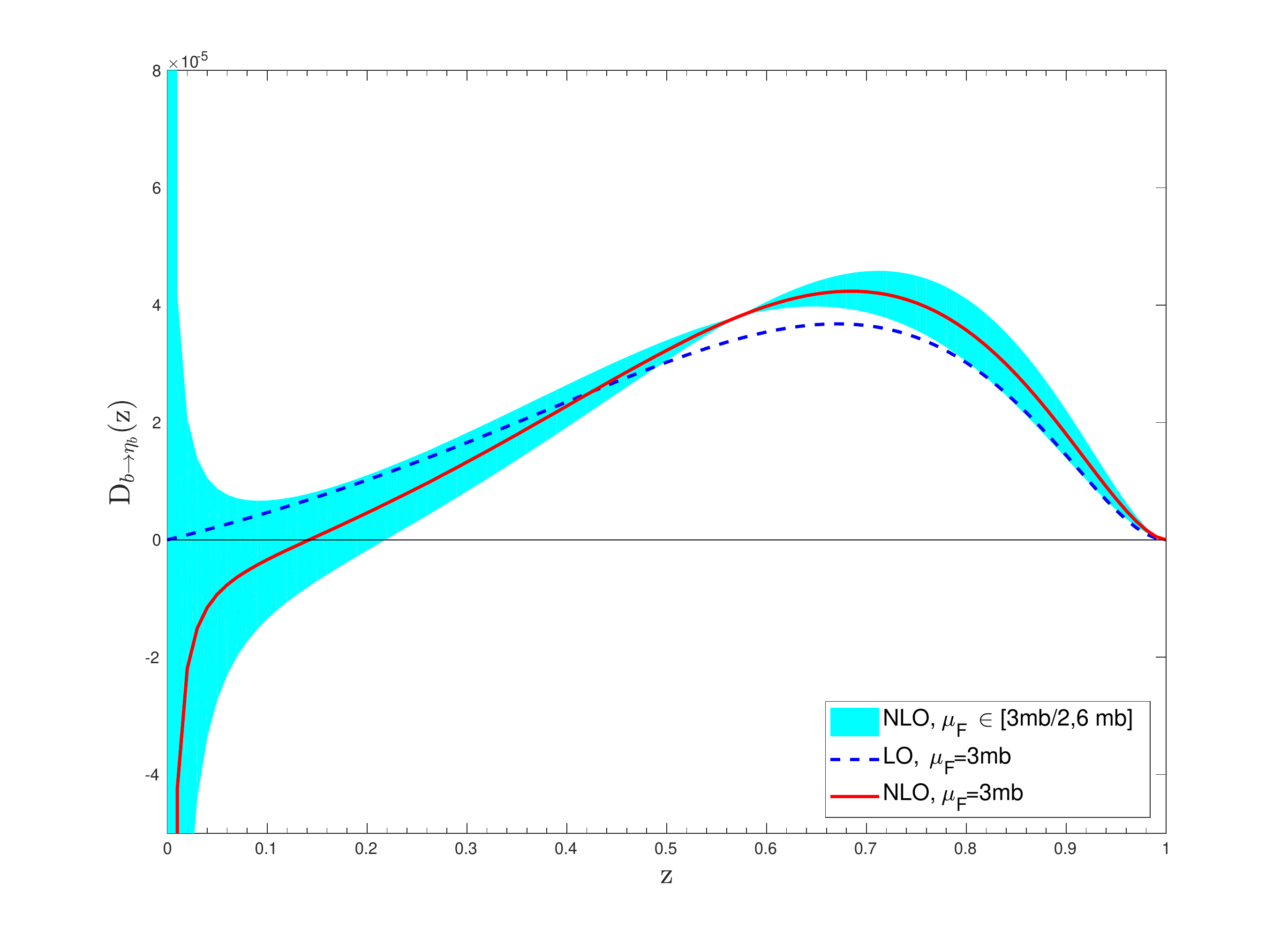}
\caption{The fragmentation function $D_{c\to \eta_c}(z,\mu_{F})$ as a function of $z$ up to LO and NLO accuracy, where $\mu_F=3m_b$ and $\mu_R=2m_b$. The bands are obtained by varying the factorization scale $\mu_F$ by a factor of 2.} \label{dz-etab-muf}
\end{figure}

The LO and NLO fragmentation functions for $b\to \eta_b$ are shown in Fig.\ref{dz-etab-mur}, the sensitivities of the LO and NLO fragmentation functions to $\mu_R$ and $\mu_F$ are shown in Fig.\ref{dz-etab-mur} and Fig.\ref{dz-etab-muf} respectively. The sensitivities of the fragmentation functions for $b\to \eta_b$ to $\mu_R$ and $\mu_F$ are similar to the $c\to \eta_c$ case.

Since the NLO fragmentation functions for $c\to \eta_c$ and $b\to \eta_b$ behave like $1/z$ as $z \to 0$, the fragmentation probabilities for $c\to \eta_c$ and $b\to \eta_b$ at the NLO level are infinite. However, the physical cross sections shall be finite, because the phase space limitations present a lower bound for $z$.

\begin{table}[h]
\centering
\begin{tabular}{|c c c c c c|}
\hline
~& n=2 & n=3 & n=4 & n=5 & n=6 \\
\hline
$M_n^{\rm LO}\times 10^5$ & ~10.5 &  ~6.78 & ~4.71 & ~3.43 & ~2.60\\
$M_n^{\rm NLO}\times 10^5$  & ~12.3 &  ~8.33 & ~5.93 & ~4.40 & ~3.37 \\
$M_n^{\rm NLO}/M_n^{\rm LO}$ & ~1.17 & ~1.23 & ~1.26 & ~1.28 & ~1.30 \\
\hline
\end{tabular}
\caption{Moments and $K$-factors for the fragmentation function $D_{c\to \eta_c}(z,\mu_F)$ with $\mu_F=3m_c$ and $\mu_R=2m_c$. }
\label{tb.moments-c}
\end{table}

\begin{table}[h]
\centering
\begin{tabular}{|c c c c c c|}
\hline
~& n=2 & n=3 & n=4 & n=5 & n=6 \\
\hline
$M_n^{\rm LO}\times 10^6$ & ~11.7 &  ~7.54 & ~5.24 & ~3.82 & ~2.89\\
$M_n^{\rm NLO}\times 10^6$  & ~12.8 &  ~8.61 & ~6.09 & ~4.50 & ~3.44 \\
$M_n^{\rm NLO}/M_n^{\rm LO}$ & ~1.09 & ~1.14 & ~1.16 & ~1.18 & ~1.19 \\
\hline
\end{tabular}
\caption{Moments and $K$-factors for the fragmentation function $D_{b\to \eta_b}(z,\mu_F)$ with $\mu_F=3m_b$ and $\mu_R=2m_b$.}
\label{tb.moments-b}
\end{table}

To see more about the effect of the NLO corrections, we calculate the moments of the fragmentation functions. The moments of the fragmentation functions $D_{Q \to \eta_Q}(z,\mu_F)$ can be defined as
\begin{eqnarray}
M_n=\int_0^1 z^{n-1} D_{Q \to \eta_Q}(z,\mu_F) dz.
\end{eqnarray}
Several moments for $c\to \eta_c$ and $b\to \eta_b$ are given in Tables \ref{tb.moments-c} and \ref{tb.moments-b}, where $K_n=M_n^{\rm NLO}/M_n^{\rm LO}$ for the $n_{\rm th}$-moment. Form those two tables, we can see that the $K$-factors of those moments are moderate although the $K$-factors of the fragmentation probabilities are divergent.

For future applications, we give fitting functions to the
NLO fragmentation functions. The NLO fragmentation functions can be written as following form
\begin{eqnarray}
D^{\rm NLO}_{Q\to \eta_Q}(z,\mu_{F})=&& D^{\rm LO}_{Q\to \eta_Q}(z)\left(1+\frac{\alpha_s(\mu_R)}{2\pi}\beta_0 {\rm ln}\frac{\mu_R^2}{4m_Q^2}\right)+\frac{\alpha_s(\mu_R)}{2\pi}{\rm ln}\frac{\mu_{F}^2}{9m_Q^2}\nonumber \\
&&\int_z^1 \frac{dy}{y}\Big[P_{QQ}(y) D_{Q\to \eta_Q}^{\rm LO}(z/y)+P_{gQ}(y) D_{g\to \eta_Q}^{\rm LO}(z/y)\Big]\nonumber \\
&& +\frac{\alpha_s(\mu_R)^3 \vert R_S(0) \vert^2}{m_Q^3} f(z).
\label{eqDzfit}
\end{eqnarray}
For $c\to \eta_c$, we have
\begin{eqnarray}
f(z)=&&0.60361\,z^8-3.48697\,z^7+8.69232 \,z^6-11.43404\, z^5 + 8.39029\, z^4 - 3.73948\, z^3\nonumber \\
&&+1.14322\, z^2 - 0.16352\, z - 0.00453-0.001282/z.
\label{eqfzfit1}
\end{eqnarray}
For $b\to \eta_b$, we have
\begin{eqnarray}
f(z)=&&1.16869\,z^8-6.27381\,z^7+14.08792 \,z^6-16.85613\, z^5 + 11.51566\, z^4 - 4.78788\, z^3\nonumber \\
&&+1.33076\, z^2 - 0.18076\, z - 0.00359-0.001282/z.
\label{eqfzfit2}
\end{eqnarray}
The difference between the $f(z)$ functions for $c\to \eta_c$ and $b\to \eta_b$ arises from the heavy quark loop in the gluon vacuum polarization. We only consider one heavy flavor (i.e., $c$) contributing to the gluon vacuum polarization for $c\to \eta_c$, while we consider two heavy flavors (i.e., $c$ and $b$) contributing to the gluon vacuum polarization for $b\to \eta_b$.

\section{Summary}

In the present paper, we have calculated the fragmentation functions for $c\to \eta_c$ and $b\to \eta_b$ up to NLO QCD accuracy. The results obtained in this paper are complementary to the previous works on the fragmentation functions for $q \to \eta_Q$ and $g\to \eta_Q$ at order $\alpha_s^3$.

The most difficult part in this work is the calculation of the real corrections. We adopt the subtraction method to calculate the real corrections. The construction of the subtraction terms and the parametrization of the phase space have been developed in our previous works.

The fragmentation functions $D_{c\to \eta_c}(z,\mu_{F})$ and $D_{b\to \eta_b}(z,\mu_{F})$ with $\mu_F=3m_Q$ and $\mu_R=2m_Q$ under the $\overline{\rm MS}$ factorization scheme are presented in the forms of figure and fitting function. The results show that the effect of the NLO corrections is significant. The fragmentation functions are decreased at small $z$ values and increased at large $z$ values after including the NLO corrections. Moreover, the NLO fragmentation functions have a singularity at $z=0$, while the LO fragmentation functions are zero at $z=0$. The sensitives of these fragmentation functions to the renormalization scale $\mu_R$ and the factorization scale $\mu_F$ are analyzed explicitly. The NLO fragmentation functions obtained in this paper can be applied to the precision predictions of the $\eta_c$ and $\eta_b$ production at high-energy colliders.

\acknowledgments
This work was supported in part by the Natural Science Foundation of China under Grants No. 11625520, No. 12005028 and No. 12047564, by the Fundamental Research Funds for the Central Universities under Grant No.2020CQJQY-Z003, and by the Chongqing Graduate Research and Innovation Foundation under Grant No.ydstd1912. \\

\providecommand{\href}[2]{#2}\begingroup\raggedright


\begin{thebibliography}{10}

\bibitem{nrqcd}
G.T. Bodwin, E. Braaten and G.P. Lepage,
\textit{Rigorous QCD analysis of inclusive annihilation and production of heavy quarkonium},
\textit{Phys. Rev}. D {\bf 51}, 1125 (1995) [\textit{Erratum-ibid.} D {\bf 55}, 5853 (1997)].

\bibitem{Brambilla:2010cs}
N.~Brambilla, S.~Eidelman, B.~K.~Heltsley, R.~Vogt, G.~T.~Bodwin, E.~Eichten, A.~D.~Frawley, A.~B.~Meyer, R.~E.~Mitchell and V.~Papadimitriou, \textit{et al.}
\textit{Heavy Quarkonium: Progress, Puzzles, and Opportunities},
\textit{Eur. Phys. J}. C \textbf{71}, 1534 (2011).

\bibitem{Brambilla:2004wf}
N.~Brambilla \textit{et al.} [Quarkonium Working Group],
\textit{Heavy quarkonium physics},
arXiv:hep-ph/0412158.

\bibitem{Butenschoen:2012px}
M.~Butenschoen and B.~A.~Kniehl,
\textit{J/psi polarization at Tevatron and LHC: Nonrelativistic-QCD factorization at the crossroads},
\textit{Phys. Rev. Lett.} \textbf{108}, 172002 (2012).

\bibitem{Chao:2012iv}
K.~T.~Chao, Y.~Q.~Ma, H.~S.~Shao, K.~Wang and Y.~J.~Zhang,
\textit{$J/\psi$ Polarization at Hadron Colliders in Nonrelativistic QCD},
\textit{Phys. Rev. Lett.} \textbf{108}, 242004 (2012).

\bibitem{Gong:2012ug}
B.~Gong, L.~P.~Wan, J.~X.~Wang and H.~F.~Zhang,
\textit{Polarization for Prompt J/\ensuremath{\psi} and \ensuremath{\psi}(2s) Production at the Tevatron and LHC},
\textit{Phys. Rev. Lett.} \textbf{110}, 042002 (2013).

\bibitem{Butenschoen:2011yh}
M.~Butenschoen and B.~A.~Kniehl,
\textit{World data of J/psi production consolidate NRQCD factorization at NLO},
\textit{Phys. Rev.} D \textbf{84}, 051501 (2011).

\bibitem{Bodwin:2014gia}
G.~T.~Bodwin, H.~S.~Chung, U.~R.~Kim and J.~Lee,
\textit{Fragmentation contributions to $J/\psi$ production at the Tevatron and the LHC},
\textit{Phys. Rev. Lett.} \textbf{113}, no.2, 022001 (2014).

\bibitem{Bodwin:2015iua}
G.~T.~Bodwin, K.~T.~Chao, H.~S.~Chung, U.~R.~Kim, J.~Lee and Y.~Q.~Ma,
\textit{Fragmentation contributions to hadroproduction of prompt $J/\psi$, $\chi_{cJ}$, and $\psi(2S)$ states},
\textit{Phys. Rev.} D \textbf{93}, no.3, 034041 (2016).

\bibitem{Collins:1989gx}
J.~C.~Collins, D.~E.~Soper and G.~F.~Sterman,
\textit{Factorization of Hard Processes in QCD},
\textit{Adv. Ser. Direct. High Energy Phys.} \textbf{5}, 1-91 (1989).

\bibitem{Kang:2011zza}
Z.~B.~Kang, J.~W.~Qiu and G.~Sterman,
\textit{Factorization and quarkonium production},
\textit{Nucl. Phys. B Proc. Suppl.} \textbf{214}, 39-43 (2011).

\bibitem{Kang:2011mg}
Z.~B.~Kang, J.~W.~Qiu and G.~Sterman,
\textit{Heavy quarkonium production and polarization},
\textit{Phys. Rev. Lett.} \textbf{108}, 102002 (2012).

\bibitem{Fleming:2012wy}
S.~Fleming, A.~K.~Leibovich, T.~Mehen and I.~Z.~Rothstein,
\textit{The Systematics of Quarkonium Production at the LHC and Double Parton Fragmentation},
\textit{Phys. Rev.} D \textbf{86}, 094012 (2012).

\bibitem{Fleming:2013qu}
S.~Fleming, A.~K.~Leibovich, T.~Mehen and I.~Z.~Rothstein,
\textit{Anomalous dimensions of the double parton fragmentation functions},
\textit{Phys. Rev.} D \textbf{87}, 074022 (2013).

\bibitem{Chang:1992bb}
C.~H.~Chang and Y.~Q.~Chen,
\textit{The Production of B(c) or anti-B(c) meson associated with two heavy quark jets in Z0 boson decay},
\textit{Phys. Rev.} D \textbf{46}, 3845 (1992),
[\textit{Erratum: Phys. Rev.} D \textbf{50}, 6013 (1994)].

\bibitem{Braaten:1993jn}
E.~Braaten, K.~m.~Cheung and T.~C.~Yuan,
\textit{Perturbative QCD fragmentation functions for $B_c$ and $B_{c}$* production},
\textit{Phys. Rev.} D \textbf{48}, 5049 (1993).

\bibitem{Braaten:1993mp}
E.~Braaten, K.~Cheung and T.~C.~Yuan,
\textit{Z0 decay into charmonium via charm quark fragmentation},
\textit{Phys. Rev.} D \textbf{48}, 4230-4235 (1993).

\bibitem{Braaten:1993rw}
E.~Braaten and T.~C.~Yuan,
\textit{Gluon fragmentation into heavy quarkonium},
\textit{Phys. Rev. Lett.} \textbf{71}, 1673-1676 (1993).

\bibitem{Braaten:1994kd}
E.~Braaten and T.~C.~Yuan,
\textit{Gluon fragmentation into P wave heavy quarkonium},
\textit{Phys. Rev.} D \textbf{50}, 3176-3180 (1994).

\bibitem{Braaten:1995cj}
E.~Braaten and T.~C.~Yuan,
\textit{Gluon fragmentation into spin triplet S wave quarkonium},
\textit{Phys. Rev.} D \textbf{52}, 6627-6629 (1995).

\bibitem{Chen:1993ii}
Y.~Q.~Chen,
\textit{Perturbative QCD predictions for the fragmentation functions of the P wave mesons with two heavy quarks},
\textit{Phys. Rev.} D \textbf{48}, 5181-5189 (1993).

\bibitem{Yuan:1994hn}
T.~C.~Yuan,
\textit{Perturbative QCD fragmentation functions for production of P wave mesons with charm and beauty},
\textit{Phys. Rev.} D \textbf{50}, 5664-5675 (1994).

\bibitem{Ma:1994zt}
J.~P.~Ma,
\textit{Calculating fragmentation functions from definitions},
\textit{Phys. Lett.} B \textbf{332}, 398-404 (1994).

\bibitem{Ma:1995ci}
J.~P.~Ma,
\textit{Gluon fragmentation into P wave triplet quarkonium},
\textit{Nucl. Phys.} B \textbf{447}, 405-424 (1995).

\bibitem{Ma:1995vi}
J.~P.~Ma,
\textit{Quark fragmentation into p wave triplet quarkonium},
\textit{Phys. Rev.} D \textbf{53}, 1185-1190 (1996).

\bibitem{Cho:1994gb}
P.~L.~Cho, M.~B.~Wise and S.~P.~Trivedi,
\textit{Gluon fragmentation into polarized charmonium},
\textit{Phys. Rev.} D \textbf{51}, 2039-2043 (1995).

\bibitem{Beneke:1995yb}
M.~Beneke and I.~Z.~Rothstein,
\textit{Psi-prime polarization as a test of color octet quarkonium production},
\textit{Phys. Lett.} B \textbf{372}, 157-164 (1996),
[\textit{Erratum: Phys. Lett.} B \textbf{389}, 769 (1996)].

\bibitem{Braaten:2000pc}
E.~Braaten and J.~Lee,
\textit{Next-to-leading order calculation of the color octet 3S(1) gluon fragmentation function for heavy quarkonium},
\textit{Nucl. Phys.} B \textbf{586}, 427-439 (2000).

\bibitem{Sang:2009zz}
W.~l.~Sang, L.~f.~Yang and Y.~q.~Chen,
\textit{Relativistic corrections to heavy quark fragmentation to S-wave heavy mesons},
\textit{Phys. Rev.} D \textbf{80}, 014013 (2009).

\bibitem{Hao:2009fa}
G.~Hao, Y.~Zuo and C.~F.~Qiao,
\textit{The Fragmentation Function of Gluon Splitting into P-wave Spin-singlet Heavy Quarkonium},
[arXiv:0911.5539 [hep-ph]].

\bibitem{Jia:2012qx}
Y.~Jia, W.~L.~Sang and J.~Xu,
\textit{Inclusive} $h_c$ \textit{Production at $B$ Factories},
\textit{Phys. Rev.} D \textbf{86}, 074023 (2012).

\bibitem{Bodwin:2014bia}
G.~T.~Bodwin, H.~S.~Chung, U.~R.~Kim and J.~Lee,
\textit{Quark fragmentation into spin-triplet S-wave quarkonium},
\textit{Phys. Rev.} D \textbf{91}, 074013 (2015).

\bibitem{Ma:2013yla}
Y.~Q.~Ma, J.~W.~Qiu and H.~Zhang,
\textit{Heavy quarkonium fragmentation functions from a heavy quark pair. I. S wave},
\textit{Phys. Rev.} D \textbf{89}, 094029 (2014).

\bibitem{Ma:2015yka}
Y.~Q.~Ma, J.~W.~Qiu and H.~Zhang,
\textit{Fragmentation functions of polarized heavy quarkonium},
\textit{JHEP} {\bf 06}, 021 (2015).

\bibitem{Yang:2019gga}
D.~Yang and W.~Zhang,
\textit{Relativistic corrections of the fragmentation functions for a heavy quark to $B_c$ and $B_c^{*}$},
\textit{Chin. Phys.} C \textbf{43}, 083101 (2019).

\bibitem{Artoisenet:2014lpa}
P.~Artoisenet and E.~Braaten,
\textit{Gluon fragmentation into quarkonium at next-to-leading order},
\textit{JHEP} \textbf{04}, 121 (2015).

\bibitem{Sepahvand:2017gup}
R.~Sepahvand and S.~Dadfar,
\textit{NLO corrections to $c$- and $b$-quark fragmentation into $j/\psi$ and $\gamma$},
\textit{Phys. Rev.} D \textbf{95}, 034012 (2017).

\bibitem{Artoisenet:2018dbs}
P.~Artoisenet and E.~Braaten,
\textit{Gluon fragmentation into quarkonium at next-to-leading order using FKS subtraction},
\textit{JHEP} \textbf{01}, 227 (2019).

\bibitem{Feng:2018ulg}
F.~Feng and Y.~Jia,
\textit{Next-to-leading-order QCD corrections to gluon fragmentation into ${}^1S_0^{(1,8)}$ quarkonia},
arXiv:1810.04138.

\bibitem{Zhang:2018mlo}
P.~Zhang, C.~Y.~Wang, X.~Liu, Y.~Q.~Ma, C.~Meng and K.~T.~Chao,
\textit{Semi-analytical calculation of gluon fragmentation into $^{1}$S$_{0}^{[1,8]}$ quarkonia at next-to-leading order},
\textit{JHEP} \textbf{04}, 116 (2019).

\bibitem{Zheng:2019dfk}
X.~C.~Zheng, C.~H.~Chang and X.~G.~Wu,
\textit{NLO fragmentation functions of heavy quarks into heavy quarkonia},
\textit{Phys. Rev.} D \textbf{100}, 014005 (2019).

\bibitem{Zheng:2019gnb}
X.~C.~Zheng, C.~H.~Chang, T.~F.~Feng and X.~G.~Wu,
\textit{QCD NLO fragmentation functions for c or $\bar{b}$ quark to Bc or Bc* meson and their application},
\textit{Phys. Rev.} D \textbf{100}, 034004 (2019).

\bibitem{Feng:2017cjk}
F.~Feng, S.~Ishaq, Y.~Jia and J.~Y.~Zhang,
\textit{Fragmentation function of gluon into spin-singlet P-wave quarkonium},
\textit{Phys. Rev.} D \textbf{102}, 014038 (2020).

\bibitem{Zhang:2020atv}
P.~Zhang, C.~Meng, Y.~Q.~Ma and K.~T.~Chao,
\textit{Gluon fragmentation into ${^{3}\hspace{-0.6mm}P_{J}^{[1,8]}}$ quark pair and test of NRQCD factorization at two-loop level},
arXiv:2011.04905.

\bibitem{Zheng:2021mqr}
X.~C.~Zheng, Z.~Y.~Zhang and X.~G.~Wu,
\textit{Fragmentation functions for a quark into a spin-singlet quarkonium: Different flavor case},
\textit{Phys. Rev.} D \textbf{103}, 074004 (2021).

\bibitem{Collins:1981uw}
J.~C.~Collins and D.~E.~Soper,
\textit{Parton Distribution and Decay Functions},
\textit{Nucl. Phys.} B \textbf{194}, 445-492 (1982).

\bibitem{Mertig:1990an}
R.~Mertig, M.~Bohm and A.~Denner,
\textit{FEYN CALC: Computer algebraic calculation of Feynman amplitudes},
\textit{Comput. Phys. Commun.} \textbf{64}, 345-359 (1991).

\bibitem{Shtabovenko:2016sxi}
V.~Shtabovenko, R.~Mertig and F.~Orellana,
\textit{New Developments in FeynCalc 9.0},
\textit{Comput. Phys. Commun.} \textbf{207}, 432-444 (2016).

\bibitem{Feng:2012iq}
F.~Feng,
\textit{$\tt{Apart}$: A Generalized Mathematica Apart Function},
\textit{Comput. Phys. Commun.} \textbf{183}, 2158-2164 (2012).

\bibitem{Smirnov:2008iw}
A.~V.~Smirnov,
\textit{Algorithm FIRE -- Feynman Integral REduction},
\textit{JHEP} \textbf{10}, 107 (2008).

\bibitem{Hahn:1998yk}
T.~Hahn and M.~Perez-Victoria,
\textit{Automatized one loop calculations in four-dimensions and D-dimensions},
\textit{Comput. Phys. Commun.} \textbf{118}, 153-165 (1999).

\bibitem{Korner:1991sx}
J.~G.~Korner, D.~Kreimer and K.~Schilcher,
\textit{A Practicable gamma(5) scheme in dimensional regularization},
\textit{Z. Phys.} C \textbf{54}, 503-512 (1992).

\bibitem{Beneke:1997zp}
M.~Beneke and V.~A.~Smirnov,
\textit{Asymptotic expansion of Feynman integrals near threshold},
\textit{Nucl. Phys.} B \textbf{522}, 321-344 (1998).

\bibitem{Mueller:1978xu}
A.~H.~Mueller,
\textit{Cut Vertices and their Renormalization: A Generalization of the Wilson Expansion},
\textit{Phys. Rev.} D \textbf{18}, 3705 (1978).

\bibitem{Eichten:1995ch}
E.~J.~Eichten and C.~Quigg,
\textit{Quarkonium wave functions at the origin},
\textit{Phys. Rev.} D \textbf{52}, 1726-1728 (1995).

\bibitem{Patrignani:2016xqp}
C.~Patrignani \textit{et al.} [Particle Data Group],
\textit{Review of Particle Physics},
\textit{Chin. Phys.} C \textbf{40}, 100001 (2016).

\bibitem{dglap1} Y.L. Dokshitzer,
\textit{Calculation of the Structure Functions for Deep Inelastic Scattering and $e^+ e^-$ Annihilation by Perturbation Theory in Quantum Chromodynamics},
\textit{Sov. Phys. JETP} {\bf 46}, 641 (1977);\textit{Zh.Eksp.Teor.Fiz.} {\bf 73}, 1216 (1977).

\bibitem{dglap2} V.N. Gribov and L.N. Lipatov,
\textit{Deep inelastic ep scattering in perturbation theory},
\textit{Sov. J. Nucl. Phys.} {\bf 15}, 438 (1972);\textit{Yad.Fiz.} {\bf 15}, 781 (1972).

\bibitem{dglap3} G. Altarelli and G. Parisi,
\textit{Asymptotic Freedom in Parton Language},
\textit{Nucl. Phys.} B {\bf 126}, 298 (1977).

\end{thebibliography}
\end{document}